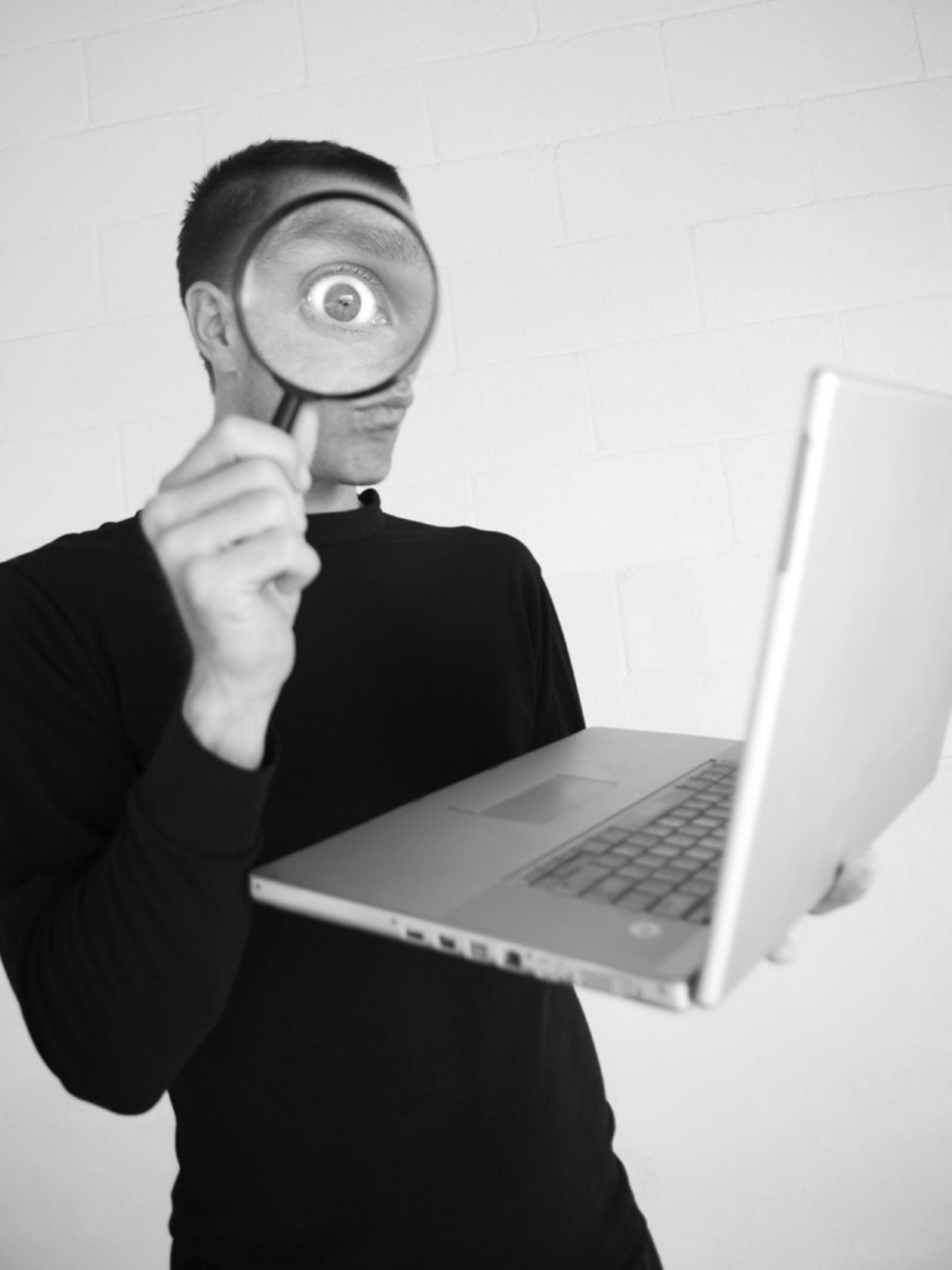

> **Frederik Zuiderveen Borgesius** pesquisador do Instituto para o Direito da Informação (IViR), Universidade de Amsterdã

# Segmentação Comportamental, Do Not Track e o desenvolvimento jurídico europeu e holandês

A segmentação comportamental (*behavioral targeting*) é o monitoramento que se faz do comportamento das pessoas na Internet ao longo do tempo, para usar as informações recolhidas com o intuito de dirigir-lhes publicidade conforme as inferências a respeito de seus interesses. Este tipo de negócio cresceu a ponto de tornar-se um mercado que movimenta milhões e milhões de dólares. Há empresas que reúnem os perfis de centenas de milhões de usuários da Internet.

As entidades reguladoras do mundo inteiro se debatem quanto a regular, ou deixar de regular, e como agir com relação à segmentação comportamental.

No dia 01 de janeiro de 2013, a versão final da Lei Holandesa das Telecomunicações entrou em vigor. Ela implementa a "cláusula dos cookies" a partir das emendas realizadas em cima da Diretiva "Privacidade e Comunicações Eletrônicas"[1]. O dispositivo holandês só permite o armazenamento e a leitura de cookies quando o usuário, uma vez informado, dá o seu consentimento. As empresas não podem inferir que houve consentimento a partir das configurações padrão de um navegador. Para facilitar a leitura, neste artigo falamos de cookies, mas o dispositivo holandês não se aplica só a eles, como também a outras tecnologias, tais como os leitores biométricos.[2]

---

[1]. Diretiva 2002/58/EC de 12 de julho de 2002 relativa ao processamento de dados pessoais e à proteção da privacidade no setor das comunicações eletrônicas (Diretiva sobre privacidade e comunicações eletrônicas) (Diário Oficial L 201, 31/07/2002 P. 0037 – 0047), conforme emendas dadas pela Diretiva 2006/24/EC [a Diretiva sobre Retenção de Dados], e a Diretiva 2009/136/EC [a Diretiva dos Direitos do Cidadão]. Este artigo usa o texto consolidado da Diretiva "Privacidade e Comunicações Eletrônicas".

A legislatura holandesa também deu um passo inédito, introduzindo uma premissa jurídica sobre o rastreamento de cookies para segmentação comportamental. Presume-se que esse uso dos cookies desencadeie o processamento de dados pessoais, onde se aplica plenamente, portanto, a lei da proteção de dados pessoais.

## :: OS COOKIES E A SEGMENTAÇÃO COMPORTAMENTAL

A segmentação comportamental é o monitoramento que se faz do comportamento das pessoas na Internet ao longo do tempo para usar as informações recolhidas com o intuito de dirigir-lhes publicidade conforme as inferências a respeito de seus interesses. Num exemplo simplificado, uma empresa pode assumir que um usuário da Internet que costuma visitar sítios sobre receitas de cozinha seja um entusiasta da gastronomia. Quando visitar um sítio de notícias, esse usuário poderá se deparar com anúncios de restaurantes ou livros de receitas. Ao visitar o mesmo sítio de notícias, alguém que leia muitos blogs de assuntos jurídicos pode se deparar com anúncios de livros de direito.

A segmentação comportamental pode beneficiar empresas e consumidores, mas também traz à baila preocupações acerca da privacidade. As empresas podem compilar perfis detalhados dos usuários da Internet com base no que eles lêem, que vídeos assistem, que buscas fazem etc.

Para realizar a segmentação comportamental, diversas tecnologias podem ser usadas – entre elas, os cookies, utilizados por muitas empresas.

O cookie é um pequeno arquivo de texto que um editor de sítios na Internet armazena no computador ou no smartphone de um usuário para reconhecer aquele dispositivo. Esses editores usam cookies para, por exemplo, lembrar o conteúdo de um carrinho de compras ("cookies de usuário" ou *first party cookies*"). Normalmente, esses cookies são relativos à "sessão", pois desaparecem depois que o usuário fecha o navegador. As empresas que atuam na segmentação comportamental costumam usar cookies persistentes para reconhecer os usuários em momentos futuros. Aquelas que publicam anúncios num sítio, como as redes de propaganda, podem colocar e ler também esses cookies persistentes ("cookies de terceiros" ou "*third party cookies*"). Resulta daí que uma rede de propaganda pode acompanhar o comportamento de um usuário da Internet em todos os sítios nos quais ela publica anúncios. Este artigo se refere a cookies que são usados para segmentação comportamental, como cookies de rastreamento.

---

2. O texto original em holandês pode ser encontrado em: http://wetten.overheid.nl/BWBR0009950.

:: A DISCUSSÃO EUROPEIA SOBRE A REGULAÇÃO DE COOKIES

Esta seção faz um apanhado geral da discussão europeia sobre a regulação de cookies, concentrando-se na Diretiva "Privacidade e Comunicações Eletrônicas". Essa lei da União Europeia de 2002 trata da proteção da privacidade no setor das comunicações eletrônicas. O Artigo 5.3 regula o uso de cookies, escutas web (*web bugs*), identificadores ocultos e outras tecnologias semelhantes[3]. As primeiras propostas para a versão de 2002 do Artigo 5.3 exigiam que as empresas pedissem consentimento antes de colocar em ação certos tipos de cookie. Depois do lobby feito junto à indústria de marketing on-line, a versão final do Artigo 5.3 usou palavreado ambíguo sobre um "direito à recusa"[4]. Costuma-se interpretar a versão de 2002 do Artigo 5.3 como uma exigência que as empresas ofereçam às pessoas a possibilidade de se opor ao uso de cookies (um sistema de escolhas livres). A versão de 2002 não deixava de exigir que as empresas informassem aos usuários qual era o uso dos cookies.

Em 2009, a Diretiva "Direitos dos Cidadãos"[5] fez emendas à Diretiva "Privacidade e Comunicações Eletrônicas". O novo Artigo 5.3 só permite o armazenamento e a leitura de cookies depois que o usuário, informado, dá o seu consentimento (note-se que certos cookies estão isentos da exigência do consentimento[6]). Em suma, o Artigo 5.3 requer um consentimento prévio por parte do usuário. Entretanto, uma frase do considerando 66 da Diretiva "Direitos do Cidadão" gerou muita confusão e discussão:

> *Onde for tecnicamente possível e eficaz, conforme os dispositivos pertinentes [da Diretiva "Proteção de Dados Pessoais"], o consentimento do usuário para o procedimento pode ser expresso através das configurações apropriadas de um navegador ou de outro aplicativo.*

A maior parte dos navegadores oferece ao usuário a possibilidade de bloquear os cookies de usuário, os cookies de terceiros, ou todos os cookies. Algumas pessoas concluem que, a partir do considerando 66, as configurações padrão do navegador podem expressar consentimento. Outras dizem que é possível inferir o consentimento para instalação de cookies a partir das configurações do navegador. Essa divergência de opiniões causou muitos debates na Europa.

---

3. Ver considerandos 24 e 25 da Diretiva "Privacidade e Comunicações Eletrônicas", e os considerandos 65 e 66 da Diretiva "Direitos do Cidadão" 2009/136. 4. S.M. Kierkegaard, 'Lobbyism and the 'opt in'/'opt out' cookie controversy. How the cookies (almost) crumbled: privacy & lobbyism' (2010) Computer Law & Security Report 2005-21, p. 310-322. 5. Diretiva 2009/136 de 25 de novembro de 2009. 6. O Artigo 5.3 da versão final da Diretiva "Privacidade e Comunicações Eletrônicas" diz: "Os EstadosMembros asseguram que o armazenamento de informações ou a possibilidade de acesso a informações já armazenadas no equipamento terminal de um assinante ou utilizador só sejam permitidos se este tiver dado o seu consentimento prévio com base em informações claras e completas, nos termos da Diretiva 95/46/CE [Data Protection Directive], nomeadamente sobre os objetivos do processamento. Tal não impede o armazenamento técnico ou o acesso que tenha como única finalidade efetuar a transmissão de uma comunicação através de uma rede de comunicações eletrônicas, ou que seja estritamente necessário ao fornecedor para fornecer um serviço da sociedade da informação que tenha sido expressamente solicitado pelo assinante ou pelo utilizador."

## :: ARGUMENTOS CONTRA O USO DAS CONFIGURAÇÕES DOS NAVEGADORES ATUAIS COMO UM MECANISMO DE CONSENTIMENTO

Alguns argumentos foram apresentados para que sejam rejeitadas as configurações do navegador como mecanismo de consentimento. O considerando 66 da Diretiva "Direitos do Cidadão" diz que, de fato, "o consentimento do usuário, que autoriza o processamento, pode ser expresso usando-se as configurações apropriadas de um navegador ou de outro aplicativo." Mas o considerando acrescenta que o consentimento precisa ser dado "em conformidade com os dispositivos pertinentes da [Diretiva "Proteção de Dados Pessoais"]". A palavra "consentimento" está definida na Diretiva "Proteção de Dados Pessoais"[7]. O consentimento deve ser (i) informado, (ii) específico, (iii) dado de livre e espontânea vontade, e (iv) a indicação do desejo de uma pessoa. Essas quatro exigências fornecem argumentos contra o uso das configurações de um navegador como mecanismo de consentimento, conforme explica-se abaixo.

Primeiro, o consentimento deve ser informado. As informações relevantes não devem ficar ocultas do usuário numa política de privacidade. Se os usuários aceitam grandes quantidades de cookies simplesmente porque não mudam as configurações de seu navegador, esse fato não pode ser considerado um consentimento informado.

Segundo, o consentimento precisa ser específico. Por exemplo, o consentimento para que dados pessoais sejam usados com propósitos comerciais não deve ser aceito. Alguns navegadores aceitam todos os cookies como padrão, inclusive os de rastreamento. Se um navegador aceita todos os cookies de usuário ou todos os cookies de terceiros, a escolha do usuário não é específica.

Terceiro, é preciso que as pessoas expressem sua vontade de dar consentimento. A não-ação do usuário dificilmente poderá ser considerada um consentimento. Além disso, as pessoas precisam ter ciência de que estão expressando uma vontade. Em princípio, a expressão de uma vontade pode ser dada de qualquer forma, e também pode ser dada de maneira implícita. Mas é improvável que todas as pessoas que *deixam* de ajustar seus navegadores com esse propósito estejam dando consentimento para todas as formas de cookies. O Tribunal de Justiça da União Europeia confirma que chegar à conclusão de que houve o consentimento não é algo trivial[8]. Portanto, as configurações de um navegador provavelmente não poderão atender às exigências de consentimento explicitadas na Diretiva "Proteção de Dados Pessoais".

Quarto, o consentimento deve ser dado de livre e espontânea vontade; portanto, o consentimento dado sob pressão não é válido. Se o usuário ajusta seu navegador para rejeitar todos os cookies, não conseguirá usar muitos dos serviços da Internet.

---

[7]. Artigo 2(h) da Diretiva de Proteção de Dados (Diretiva 95/46/EC de 24 de outubro de 1995). O considerando 17 da Diretiva "Privacidade e Comunicações Eletrônicas" se refere à Diretiva de Proteção de Dados para a definição do consentimento. [8]. ECJ: Case C-92/09 and C-93/09 *Volker und Markus Schecke GbR* [2010], para 63. ECJ: Case C112/11, ebookers.com [2012].

Entretanto, se o navegador for configurado para aceitar os cookies de usuário, também aceitará cookies de rastreamento. Além disso, algumas empresas levam os navegadores que aceitam cookies de usuário a aceitar também os cookies de rastreamento[9]. Em suma, não dá para ter certeza se os usuários realmente têm a liberdade de configurar seus navegadores para rejeitar todos os cookies.[10]

É importante ressaltar que o Tribunal de Justiça da União Europeia diz que a Diretiva "Privacidade e Comunicações Eletrônicas" deve ser interpretada juntamente com os direitos fundamentais.[11] A Diretiva "Privacidade e Comunicações Eletrônicas" visa proteger o direito à privacidade, o direito à proteção dos dados pessoais e a confidencialidade das comunicações. Todos esses direitos estão incluídos na Carta dos Direitos Fundamentais da União Europeia. O considerando 24 do preâmbulo da Diretiva "Privacidade e Comunicações Eletrônicas" diz que os aparelhos dos usuários "fazem parte da esfera privada de usuários, que requerem proteção conforme a Convenção Europeia para a Proteção dos Direitos Humanos e as Liberdades Fundamentais." O Tribunal Europeu dos Direitos Humanos interpreta o direito ao amplo respeito pela vida privada. Além disso, a Carta e outros Tratados da União Europeia enfatizam a importância de garantir-se um elevado nível de proteção ao consumidor[12]. Em suma, as leis na União Europeia requerem uma interpretação da Diretiva "Privacidade e Comunicações Eletrônicas" pela ótica da privacidade.

Por outro lado, a proteção de dados pessoais e o direito à privacidade não são absolutos. Os Estados-membros devem atingir um equilíbrio justo entre os direitos fundamentais ao aplicar as diretivas. As empresas que empregam cookies também poderiam invocar um direito fundamental da Carta: "a liberdade de tocar uma empresa em conformidade com o direito da União e as leis e práticas nacionais" [13]- embora seja pouco plausível que a liberdade de tocar uma empresa implique no direito de armazenar cookies nos aparelhos das pessoas sem o consentimento adequado. Em suma, aceitar as configurações padrão do navegador como consentimento para o uso de cookies é um tanto difícil de conciliar com a lei na União Europeia.

:: ARGUMENTOS A FAVOR DO USO DAS CONFIGURAÇÕES DOS NAVEGADORES ATUAIS COMO UM MECANISMO DE CONSENTIMENTO

Há quem seja a favor do uso das configurações padrão dos navegadores como um mecanismo de consentimento. Os argumentos podem ser resumidos assim: primeiro, as pessoas que não mudam as configurações de seus navegadores consentem implicitamente com o armazenamento de cookies em seus computadores. O *Interactive Advertising Bureau* do Reino Unido diz, por exemplo: "Acreditamos que as configurações padrão do navegador possam ser equivalentes a 'consentimento' conforme disposto em nosso considerando 66" (ênfase original).[14]

---

9. J. Mayer, Safari Trackers (Web Policy, 17 de fevereiro de 2012), http://webpolicy.org/2012/02/17/safari-trackers; B. Krishnamurthy & C. Wills, 'Privacy diffusion on the web: a longitudinal perspective' (2009) WWW '09: Proceedings of the 18th international conference on World wide web, ACM, 2009, p. 544.   10. Grupo de Trabalho do Artigo 29, 'Opinião 2/2010 sobre publicidade baseada em segmentação comportamental' (WP 171, 22 June 2010); Grupo de Trabalho do Artigo 29, 'Opinião 15/2011 sobre a definição de consentimento' (WP 187, 13 July 2011). O Grupo de Trabalho do Artigo 29 foi instituído pelo artigo 29.o da Diretiva 95/46/CE. Trata-se de um órgão consultivo europeu independente em matéria de proteção dos dados e da privacidade. As suas atribuições estão descritas no artigo 30.o da Diretiva 95/46/CE e no artigo 15.o da Diretiva 2002/58/CE.   11. ECJ: Caso C-275/06, *Promusicae* [2008], par. 67-68, and dictum. Ver também o considerando 62 da Diretiva de Direitos dos Cidadãos 2009/136.   12. Ver artigos 38 e 51(1) da Carta dos Direitos Fundamentais da União Europeia, e artigos 12, 114(3) e 169 do Tratado sobre o Funcionamento da União Europeia.   13. Artigo 16 da Carta de Direitos Fundamentais da União Europeia.

Segundo, há quem tema que os usuários da Internet sejam bombardeados com pop-ups pedindo consentimento se as configurações do navegador não forem suficientes para sinalizar consentimento. Assim, navegar pela Internet seria um pesadelo. Ficaria difícil conciliar isso com o considerando 66 da Diretiva "Direitos do Cidadão", que pede uma solução de fácil uso para o usuário: "Os métodos de fornecer informação e oferecer o direito a recusar devem ser o mais fácil possível para o usuário."

Terceiro, os pop-ups talvez não consigam passar a informação de maneira significativa para o usuário. Muitos são os que clicam em "Concordo" com qualquer declaração que lhes seja apresentada. Esta situação pode piorar se as pessoas começarem a ver mais pop-ups aparecendo em suas janelas. Portanto, a lei não seria eficaz no sentido de proteger a privacidade dos usuários. Há até quem argumente que os usuários podem clicar distraidamente num *spyware* ou num vírus se

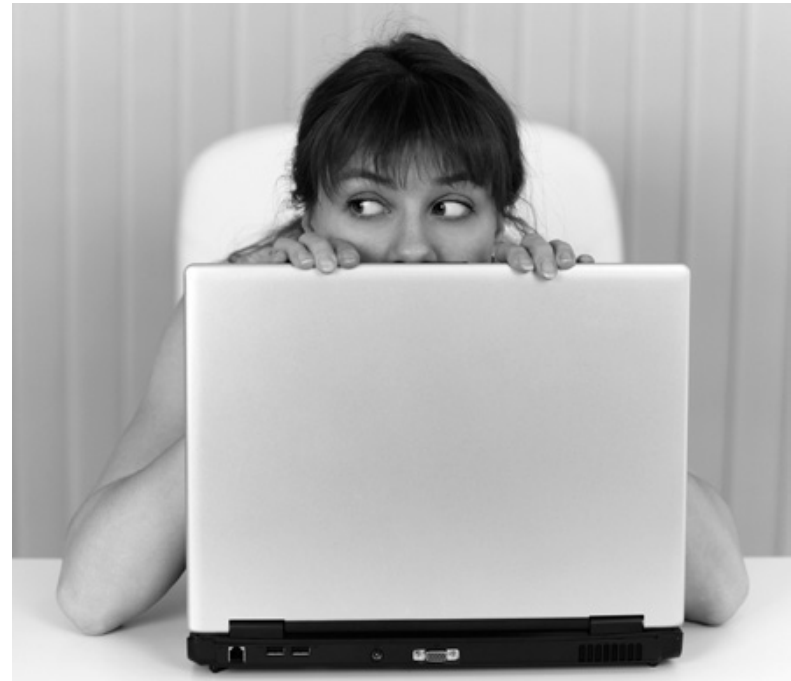

▎Muitos são os que clicam em "Concordo" com qualquer declaração que lhes seja apresentada.

passarem a ter de aceitar muitos pop-ups. Por outro lado: se as configurações do navegador expressam consentimento, isso pode implicar, para terceiros, na premissa de que os usuários consentem com a instalação de *spyware* uma vez que seus navegadores não bloqueiam *spyware*.[15] Ainda assim, é preciso

---

14. Interactive Advertising Bureau, Resposta da IAB UK à consulta do Department for Business, Innovation & Skills sobre a implementação da revisão do marco das comunicações eletrônicas na UE (IAB 1 de dezembro de 2010, www.iabuk.net/sites/default/files/IABUKresponsetoBISconsultationonimplementingthe-revisedEUElectronicCommunicationsFramework_7427_0.pdf), p. 2.   15. N.A.N.M. Van Eijk e outros, A bite too big: Dilemma's bij de implementatie van de Cookiewet in Nederland, TNO-report no. 35473. http://www.ivir.nl/publicaties/vaneijk/A_bite_too_big.pdf, p. 63.

verificar se os pop-ups vão conseguir informar direito os usuários.

Quarto, as pessoas podem acabar clicando em "não" para o rastreamento de cookies usados para fazer segmentação comportamental quando se depararem com um pedido de consentimento. Isso reduziria a receita oriunda de anúncios, e poderia acabar com a oferta gratuita de alguns serviços. Outro argumento pertinente é que uma regulamentação sobre cookies forte demais é ruim para a competitividade dos provedores de conteúdos europeus na Internet e outras empresas que utilizem cookies. Por exemplo, os sítios norte-americanos podem ignorar a regra da União Europeia.

Quinto, a lei não explica como as empresas devem obter o consentimento. Há quem diga que não está clara a maneira como as empresas devem cumprir a lei, se não quiserem usar os pop-ups. Isso seria prejudicial para a segurança jurídica.

Em suma, há argumentos para duvidar da razoabilidade da regra do consentimento para cookies na Diretiva "Privacidade e Comunicações Eletrônicas". Alguns desses argumentos são usados a favor da aceitação das configurações padrão do navegador como mecanismo de consentimento. É difícil avaliar o argumento de que o rastreamento dos cookies é necessário para a publicidade que financia os serviços de Internet, afinal também pode haver anúncios que não dependam do monitoramento do comportamento das pessoas na Internet. Além disso, é rara a pesquisa independente em torno da renda gerada a partir da segmentação comportamental[16]. E mais, não parece plausível que a importância da renda oriunda de publicidade venha a ditar a interpretação de uma diretiva.

O temor de que surja uma avalanche de pop-ups parece ser o argumento mais convincente em favor da aceitação das configurações dos navegadores atuais como um mecanismo de consentimento. Mas esse temor pode estar algo exagerado, pois certos tipos de cookies estão isentos da exigência de consentimento. É possível construir sítios Web sem cookies que precisem de consentimento. Além disso, os usuários da Internet só precisam dar consentimento para um cookie específico uma vez. Não obstante, ficar constantemente clicando em pop-ups para desativá-los não parece nada atraente.

:: DO NOT TRACK

O Tribunal de Justiça da União Europeia tem a palavra final sobre a interpretação que se deve dar à Diretiva "Privacidade e Comunicações Eletrônicas". Não obstante, a comissária Kroes já deu sua interpretação. Ela sugere que um sistema de *Do Not Track* possibilitaria às empresas o cumprimento da Diretiva "Privacidade e Comunicações Eletrônicas"[17]. Uma opção de *Do Not Track*[18] nos navegadores deveria permitir que os usuários sinalizassem que

---

16. J. R. Mayer & J. C. Mitchell, "Third-Party Web Tracking: Policy and Technology", IEEE Security & Privacy, novembro/dezembro de 2012.
17. Neelie Kroes é Vice-Presidente da Comissão Europeia responsável pela Agenda Digital. Ver "Online privacy – reinforcing trust and confidence", comunicado de imprensa divulgado durante o Online Tracking Protection & Browsers Workshop, em junho de 2011, Bruxelas (http://europa.eu/rapid/pressReleasesAction.do?reference=SPEECH/11/461). 18. N.E.: Do Not Track é uma proposta tecnológica e política que permite ao usuário da Web optar por não aceitar o rastreamento feito por terceiras partes através de um cabeçalho no protocolo http que sinaliza sua opção. Ver em http://donottrack.us/

não querem ser rastreados a partir dos sítios que visitam na Internet.

O Consórcio World Wide Web é uma comunidade internacional onde as organizações membro cooperam para a elaboração de padrões na Internet. Desde novembro de 2011, um Grupo de Trabalho de Proteção contra o Rastreamento, formado por essa comunidade, está envolvido numa discussão sobre uma norma *Do Not Track*[19]. Basicamente, essa norma deverá permitir que as pessoas rejeitem o rastreamento.

Uma norma assim permitiria que as pessoas sinalizassem com seus navegadores que não querem ser rastreadas. Uma empresa que receba um sinal "Não me rastreiem" deve responder: "OK, não vou rastreá-lo". Mas se a empresa continuar rastreando o usuário depois de uma resposta assim, a lei deve ser acionada. Por exemplo, as autoridades responsáveis pela proteção ao consumidor devem poder interferir se uma empresa descumprir a promessa que fez ao responder "Não vou rastreá-lo".

Mas como um sistema *Do Not Track* poderia ajudar as empresas a cumprir a regra do consentimento do Artigo 5.3 da Diretiva "Privacidade e Comunicações Eletrônicas"? Talvez seja possível um acordo nos seguintes moldes: as empresas devem deixar de rastrear os usuários europeus da Internet que não configurarem uma preferência de *Do Not Track*. Se alguém sinalizar para uma empresa "Sim, pode me rastrear" depois de receber informações suficientes sobre o assunto, essa empresa poderá colocar um cookie de rastreamento. Portanto, na Europa, deixar de configurar uma preferência teria o mesmo efeito que configurar uma preferência por *Do Not Track*. Nos países onde a lei não exija consentimento, as empresas ficam livres para rastrear as pessoas que não determinarem essa preferência.

As propostas em favor de uma norma do tipo *Do Not Track* excluem o rastreamento dentro de um mesmo sítio. Portanto, a norma permitiria que empresas como a Amazon ou o Facebook analisem o comportamento das pessoas dentro de seu próprio sítio, mesmo que as pessoas acionem o sinal de *Do Not Track*. Por outro lado, a regra de consentimento da Diretiva "Privacidade e Comunicações Eletrônicas" também se aplica a cookies de rastreamento de usuário.

Um dos principais pontos de discórdia é se o *Do Not Track* significa "Bloqueio de Coleta" (não use tecnologias de rastreamento) ou simplesmente "Não use para segmentação de comportamento" (continue coletando dados mas pare de mostrar a publicidade direcionada). Outra discórdia é se devem ser respeitados os navegadores que utilizem *Do Not Track* como padrão. Algumas empresas de publicidade dizem que a adoção de *Do Not Track* como padrão pelo navegador pode ser ignorada pois ela não expressam a vontade do usuário. No momento em que escrevi estas linhas, o Grupo de Trabalho sobre a Proteção contra o Rastreamento não tinha chegado a um consenso ainda.

---

[19]. Ver a discussão na lista pública de e-mails do Grupo de Trabalho sobre Do Not Track do Consórcio World Wide Web: http://lists.w3.org/Archives/Public/public-tracking/.

Resumindo, a discussão jurídica sobre os cookies e a segmentação comportamental na Europa foi lançada em 2009 como emendas à Diretiva "Privacidade e Comunicações Eletrônicas". A discussão se concentra na maneira como as empresas devem obter o consentimento dos usuários da Internet para instalarem seus cookies. O principal ponto de discórdia é se os navegadores atuais são ajustados para indicar consentimento para o uso de cookies. Há argumentos contra e a favor da aceitação das configurações dos atuais navegadores como um mecanismo de consentimento. O debate ainda não terminou. Enquanto isso, está sendo discutido um sistema de *Do Not Track*, mas ainda não ficou clara a maneira como isso ajudaria as empresas a cumprir a Diretiva "Privacidade e Comunicações Eletrônicas".

:: ANÁLISE DA LEI HOLANDESA DE TELECOMUNICAÇÕES

Nos Países Baixos, a regra do consentimento para a Diretiva "Privacidade e Comunicações Eletrônicas" está implementada no Artigo 11.7a da Lei Holandesa de Telecomunicações. A regra geral determina que as empresas que querem armazenar ou rastrear um cookie no aparelho do usuário devem: (a) dar ao usuário informações claras e completas sobre o propósito daquele cookie, e (b) obter consentimento do usuário[20].

Ficam isentas da exigência de consentimento duas categorias de cookies funcionais. Primeiro, não é necessário consentimento para cookies cujo único propósito seja o transporte de uma comunicação através de uma rede eletrônica de comunicações. Um exemplo seriam os cookies necessários para rotear a informação dentro da rede. Segundo, não é necessário consentimento para um cookie que seja estritamente necessário para a prestação de um serviço que o usuário tenha solicitado. Por exemplo, um cookie para um carrinho de compras digital.

O elemento mais marcante da implementação holandesa é uma premissa da lei sobre os cookies de rastreamento: presume-se que o uso desses cookies leve ao processamento de dados pessoais. Portanto, aplica-se a Lei Holandesa de Proteção de Dados Pessoais. Na maioria dos casos, isso significa que a empresa lançadora do cookie deva obter consentimento prévio do usuário sem existência de ambiguidades. O ônus da prova recai sobre as empresas que empreguem os cookies de rastreamento, no sentido de provarem que não processam dados pessoais.

Assim como o Artigo 5.3 da Diretiva "Privacidade e Comunicações Eletrônicas", o Artigo 11.7 da Lei Holandesa de Telecomunicações tem um escopo amplo. Ele se aplica a qualquer "armazenamento de informação" ou "acesso a informação já armazenada" no equipamento terminal de um usuário.

---

20. Para ver uma versão em inglês da provisão: F. Zuiderveen Borgesius, Behavioral Targeting: Legal Developments in Europe and the Netherlands, paper elaborado para o workshop Do Not Track do W3C, novembro de 2012, www.ivir.nl/publications/borgesius/Position_paper_W3C.pdf, p. 5.

Esse equipamento terminal do usuário inclui, por exemplo, computadores e smartphones. O dispositivo holandês se aplica a qualquer informação armazenada ou acessada no aparelho do usuário, como cookies, *spyware* e vírus. Por exemplo, os reguladores holandeses já aplicaram o antecessor do Artigo 11.7a a um *spyware*. O histórico legislativo mostra que o Artigo 11.7a também se aplica a identificadores biométricos e a acesso a informações num decodificador de TV digital para segmentação comportamental[21].

O parágrafo 2 enfatiza que as exigências de informação e consentimento também se aplicam quando alguém instala, por outros meios que não a Internet, software capaz de armazenar ou acessar informações no aparelho de um usuário através da Internet. Um exemplo disso seria uma empresa distribuindo CDs com software que lhe dê acesso a informações no computador do usuário após a instalação. O dispositivo se aplicaria, por exemplo, aos CDs distribuídos pela SONY em 2005, que instalavam *spyware* no computador das pessoas.

## :: REGRA PRINCIPAL: CONSENTIMENTO INFORMADO

A regra geral no Artigo 11.7a da Lei de Telecomunicações basicamente copia o Artigo 5.3 da versão final da Diretiva "Privacidade e Comunicações Eletrônicas". Uma empresa que queira acessar informações armazenadas no

> ▎O ônus da prova recai sobre as empresas que empreguem os cookies de rastreamento, no sentido de provarem que não processam dados pessoais.

equipamento terminal de um usuário, ou que queira armazenar informações nesse equipamento, deverá dar ao usuário informações claras, pelo menos sobre o propósito do cookie, e deverá também obter o consentimento do mesmo. Em suma, cookies só são permitidos após consentimento informado por parte do usuário. Há, entretanto, uma discussão sobre exceções a essa regra.

As informações prestadas ao usuário devem ser "claras e completas" e devem estar "em conformidade com a Lei de Proteção de Dados Pessoais". As empresas devem, no mínimo, explicar o propósito de um cookie. A Lei de Proteção de Dados Pessoais exige que as empresas forneçam mais informações detalhadas, caso isso seja necessário para assegurar um processamento justo e legítimo dos dados.

Organizações como o *Interactive Advertising Bureau* holandês fizeram lobby em prol de uma implementação que aceitaria as configurações padrão do navegador como consentimento[22]. Os legisladores holandeses não seguiram a sugestão. O memorial explicativo das emendas à Lei das Telecomunicações diz que os navegadores atuais não são adequados para dar consentimento. Por exemplo, muitos deles aceitam cookies como padrão. Os legisladores acrescentaram que existe a possibilidade de navegadores que venham a oferecer funções para a expressão do consentimento. Eles não explicaram como as empresas devem obter o consentimento para cookies, mas convocam as empresas do ramo da propaganda on-line para encontrar uma solução viável.

---

21. Eerste Kamer, vergaderjaar 2011–2012, 32 549, G, 17 February 2012, p. 4-6.

A Agência Nacional holandesa encarregada das telecomunicações, a OPTA, diz que o consentimento pode ser obtido através de uma janela pop-up. Diz também que um anunciante na Internet pode negar acesso a visitantes que não aceitem um cookie. É possível questionar se o usuário dá consentimento "de livre e espontânea vontade" em casos assim. Por outro lado, o considerando 25 do preâmbulo da Diretiva "Privacidade e Comunicações Eletrônicas" permite que anunciantes na Internet condicionem o acesso "com base numa aceitação bem-informada acerca de um cookie ou de um dispositivo semelhante, se ele for usado para um propósito legítimo."

As empresas não precisam pedir novo consentimento para visitantes que estejam voltando ao seu sítio Web. Se um usuário dá o seu consentimento para que ela armazene um cookie permanente, o consentimento permanece válido durante toda a vida útil do cookie, conforme a OPTA. Assim, as pessoas só precisam dar consentimento uma vez a um cookie desse tipo. Com isso, minimiza-se o risco de uma avalanche de pop-ups de consentimento. Além disso, muitos cookies estão isentos da exigência de consentimento. Em suma, o dispositivo holandês somente permite o armazenamento e a leitura de cookies depois do consentimento informado do usuário, a menos que se aplique uma exceção.

:: DUAS EXCEÇÕES À EXIGÊNCIA DE CONSENTIMENTO

Assim como na Diretiva "Privacidade e Comunicações Eletrônicas", o dispositivo holandês isenta duas categorias de cookies da exigência de consentimento. Primeiro, uma empresa que queira armazenar ou ler informações "com o único propósito de transportar uma comunicação através de uma rede de comunicações eletrônicas" não precisa obter o consentimento prévio. Uma segunda exceção se refere a cookies que são "estritamente necessários" para "a prestação de um serviço da sociedade da informação solicitado pelo usuário". Um serviço da sociedade da informação significa, grosso modo: um serviço prestado via Internet[23].

---

22. Ver: Interactive Advertising Bureau, IAB Europe urges EU Member States to consider negative impact of an overly strict consent for cookies (www.iabeurope.eu/news/iab-europe-urges-eu-member-states-to-consider-negative-impact-of-an-overly-strict-consent-for-cookies.aspx). 23. O Artigo 1(2) da Diretiva 98/34/EC define assim um serviço da sociedade da informação : "qualquer serviço remunerado normalmente prestado à distância, por meios eletrônicos e mediante a solicitação individual do usuário do serviço." Este artigo está implementado no Artigo 3:15d do Código Civil Holandês.

Um exemplo seria um cookie para um carrinho de compras digital.

O Grupo de Trabalho do Artigo 29 emitiu um parecer sobre as isenções da exigência de consentimento[24]. Os pareceres do Grupo de Trabalho não têm força de lei, mas influenciam, uma vez que o Grupo de Trabalho é formado por representantes das autoridades responsáveis pela proteção de dados dos estados membros e pelo Supervisor Europeu para a Proteção de Dados, e normalmente toma decisões por consenso.

O Grupo de Trabalho usa uma interpretação estreita dos cookies que são necessários para transportar uma comunicação. Para ser isenta, a transmissão de uma comunicação através de uma rede de comunicações eletrônicas precisa ficar impossibilitada sem o cookie. O Grupo de Trabalho também dá exemplos de cookies que são estritamente necessários para um serviço de Internet solicitado pelo usuário. A saber, um cookie necessário para um carrinho de compras digital, para um procedimento de login, ou para lembrar as preferências de idioma estabelecidas pelo usuário.

O Grupo de Trabalho confirma que certos tipos de cookies não são isentos da exigência de consentimento. A exigência de consentimento prévio dado por um usuário informado se aplica ao rastreamento com cookies tais como os que são usados para a segmentação comportamental. O consentimento é exigido para outros cookies de terceiros que também são usados para a publicidade. A regra do consentimento se aplica ainda a muitos cookies que são usados pelos sítios de redes sociais para rastrear os usuários quando estes navegam pela Internet.

:: PREMISSA LEGAL ACERCA DOS COOKIES DE RASTREAMENTO

Durante o processo legislativo, o Parlamento Holandês acrescentou uma emenda relativa ao uso de cookies de rastreamento e tecnologias semelhantes. Entende-se que esse uso de cookies envolva o processamento de "dados pessoais". A premissa legal reverte o ônus da prova. Cabe às empresas que empregam os cookies de rastreamento provar que não processam dados pessoais. A Lei Holandesa parece ser a primeira do mundo que parte de uma premissa jurídica desse tipo para os cookies de rastreamento - o Artigo 11.7a traz a seguinte redação:

*Qualquer atividade mencionada no preâmbulo, com vistas a coletar, analisar ou combinar informações sobre o uso que um assinante ou usuário faz dos vários serviços da sociedade da informação, com propósitos comerciais, idealísticos ou de caridade, é tida como processamento de dados pessoais, conforme definido no Artigo 1(b) da Lei de Proteção de Dados.*

---

24. Grupo de Trabalho do Artigo 29, 'Opinião 04/2012 sobre Exceções ao Consentimento para Cookies' (WP 194, 7 de junho de 2012).

A frase "atividade mencionada no preâmbulo" refere-se a acessar informações armazenadas no equipamento terminal de um usuário ou a armazenar informações no equipamento terminal de um usuário, através de redes de comunicações eletrônicas. Um "serviço da sociedade da informação" é identificado de forma ampla. A definição cobre a maior parte dos serviços de Internet, inclusive os gratuitos. Um exemplo do "uso de vários serviços da sociedade da informação" seria a visita a variados sítios na Internet.

Portanto, se uma empresa quiser acessar ou armazenar informações para rastrear pessoas em vários serviços de Internet, a premissa jurídica se aplica. A frase "com propósitos comerciais, idealísticos ou de caridade" significa, grosso modo: com propósitos de marketing direto. O escopo da frase holandesa é um tanto mais amplo, e também inclui mensagens voltadas para angariar verbas, por exemplo. O memorial explicativo diz que a premissa legal visa assegurar que a Lei de Proteção de Dados se aplique quando uma empresa emprega cookies de rastreamento.

O memorial explicativo diz que uma empresa precisa obter um consentimento do usuário sem ambiguidade antes de colocar um cookie de rastreamento. O governo holandês mais tarde disse que, em princípio, outras bases além do consentimento sem ambiguidade poderiam legitimar o processamento de dados pessoais de natureza não sensível. Entretanto, o Grupo de Trabalho sugere que o processamento de dados pessoais para segmentação comportamental raramente pode ser legitimado por outro embasamento que não o "consentimento sem ambiguidade" do usuário. Se uma empresa processa dados pessoais, ela precisa cumprir com todos os princípios relativos à proteção de dados. Por exemplo, as empresas devem evitar a coleta de dados excessivos ou sigilosos.

A premissa legal entrou em vigor em 1 de janeiro de 2013, embora o resto do dispositivo já esteja em vigor desde junho de 2012. O Senado Holandês disse que esse retardo poderia permitir que a indústria de marketing on-line apresentasse um sistema fácil de usar para a obtenção do consentimento, por exemplo, elaborando uma norma palpável para o *Do Not Track*.

:: FISCALIZAÇÃO E EFEITO
EXTRATERRITORIAL

Quem precisa cumprir o Artigo 11.7a da Lei Holandesa de Telecomunicações? O Parágrafo 1 diz: "qualquer um" que queira acessar informações armazenadas no aparelho de um usuário, ou que queira armazenar informações no aparelho de um usuário.

O dispositivo holandês se aplica a empresas que não estejam sediadas nos Países Baixos? Este assunto é complicado. A OPTA, autoridade responsável pelas telecomunicações, supervisiona

o cumprimento da Lei Holandesa de Telecomunicações. Conforme essa entidade, o Artigo 11.7a também se aplica a anunciantes estrangeiros na Internet. O texto da Diretiva "Privacidade e Comunicações Eletrônicas" permite essa interpretação.

Outra questão a se discutir é quem deve pedir o consentimento do usuário. Se um usuário visita um sítio de notícias e vinte empresas de segmentação comportamental colocam cookies espalhados pelo sítio, quem deve pedir o consentimento do usuário? A OPTA diz que um anunciante na Internet compartilha a responsabilidade com terceiros que espalhem cookies pelo seu sítio. Isso se alinha com o conselho anterior do Grupo de Trabalho do Artigo 29.

Os reguladores holandeses não começaram a fazer cumprir a lei ainda. A OPTA tem autoridade para aplicar multas de até 450.000 euros pelo seu descumprimento. Se uma empresa processa "dados pessoais", aplica-se a Lei de Proteção de Dados. Nesse caso, uma segunda autoridade reguladora entra em cena. A Autoridade Holandesa de Proteção de Dados supervisiona o cumprimento da Lei de Proteção de Dados. A Autoridade de Proteção de Dados não precisará provar que uma empresa usando um cookie de rastreamento processa dados pessoais. Caberá à empresa provar que não o faz. A Autoridade de Proteção de Dados não pode aplicar multas pelo descumprimento do dispositivo holandês a respeito dos cookies.

Mas pode aplicar vultosas penalidades preventivas se uma empresa desobedecer suas ordens administrativas.

### :: CONCLUSÃO

Este artigo analisou as novas regras holandesas sobre cookies e tecnologias semelhantes. Primeiro, foi abordada a discussão europeia sobre a regulação da segmentação comportamental e dos cookies. Na Europa, a discussão está centrada na maneira como as empresas devem obter o consentimento dos usuários.

Nos Países Baixos, o Artigo 5.3 da Diretiva "Privacidade e Comunicações Eletrônicas" está implementado na Lei Holandesa das Telecomunicações. O dispositivo holandês, que também se aplica a empresas fora dos Países Baixos, só permite o armazenamento e leitura de cookies e arquivos assemelhados depois de o usuário emitir um consentimento informado. As empresas não podem inferir consentimento a partir das configurações padrão dos navegadores atuais. O legislador holandês acrescentou uma premissa legal sobre o rastreamento de cookies e tecnologias semelhantes. Presume-se que usar cookies dessa maneira envolva o processamento de dados pessoais. Ainda não está clara a maneira como a Lei Holandesa vai ser aplicada. Ela poderá trazer más notícias para empresas que empreguem cookies de rastreamento e tecnologias semelhantes, especialmente se outros países resolverem seguir o caminho holandês. ●